\documentclass[a4paper,10pt]{article}

\usepackage[utf8]{inputenc}
\usepackage[english]{babel}
\usepackage{amsmath}
\usepackage{amssymb}
\usepackage{graphicx}
\usepackage{bm}
\usepackage{authblk}
\usepackage{hyperref}
\usepackage[left=2cm,right=2cm,top=3cm,bottom=3cm]{geometry}

\author[1,2]{V.V.~Samsonov}
\author[2,3]{K.V.~Nikolaev}
\author[4,5]{B.I.~Ostrovskii}
\author[2]{S.N.~Yakunin}

\affil[1]{Moscow State University, Moscow, Russia}
\affil[2]{National Research Center Kurchatov Institute, Moscow, Russia}
\affil[3]{Moscow Institute of Physics and Technology, Dolgoprudny, Russia}
\affil[4]{Crystallography Institute, Kurchatov Complex Crystallography and Photonics, NRC ``Kurchatov Institute'', Moscow, Russia}
\affil[5]{Institute of Solid State Physics, Russian Academy of Sciences, Chernogolovka, Russia}

\title{\bf X-ray diffraction from smectic multilayers: crossover from kinematical to dynamical regime }

\date{\today}

\begin{document}

\maketitle

\begin{abstract}
    We study X-ray diffraction in smectic liquid crystal multilayers. 
    Such systems are fabricated as freely suspended films and have a unique layered structure. 
    As such, they can be described as organic Bragg mirrors with sub-nanometer roughness. 
    However, an interesting peculiarity arises in the diffraction on these structures: the characteristic shape of diffraction peaks associated with dynamical scattering effects is not observed.
    Instead, the diffraction can be well described kinematically, which is atypical for Bragg mirrors. 
    In this article we investigate the transition between the kinematical and dynamical regimes of diffraction. 
    For this purpose, we analyze the reflection of synchrotron radiation on a real liquid crystal sample with both kinematical and dynamical theories. 
    Furthermore, based on these theories, we derive a quantitative criterion for the transition from the kinematical to the dynamical regime. 
    This, in turn, allows us to explain the peculiar diffraction behavior in smectic films with thicknesses exceeding thousands of molecular layers.
\end{abstract}

\section{Introduction}

    High X-ray reflectivity can be achieved using a periodic structure in which weakly reflected waves interfere constructively at certain angles,
    giving rise to Bragg peaks -- so called multilayer or Bragg mirrors~\cite{attwood2000soft, louis2011nanometer}.
    These mirrors are typically composed of two inorganic materials of alternating high and low electron densities,
    designed to maximize normal-incidence reflectivity within a narrow spectral bandwidth.
    Bragg mirrors are predominantly amorphous within their planes,
    and are well adaptable to curved interfaces,
    enabling their usage in reflective optics,
    X-ray microscopy, telescopes,
    and other advanced applications~\cite{yakshin2010mirrors,pleshkov2021multilayers, spiga2006multilayer, salmaso2011multilayer, chkhalo2016multilayer}.

    Although liquid crystals (LCs) have been known for a quite long time, their layered variants have rarely been considered as potential Bragg mirrors. 
    These organic materials primarily consist of low atomic number atoms such as carbon, hydrogen, nitrogen and oxygen, which, at first glance, may prevent their use as the elements of reflective optics. 
    However, as demonstrated below, when LC molecules are arranged in highly ordered stacks comprising thousands of molecular layers, 
    their reflectance properties become comparable to those of conventional inorganic Bragg mirrors. 
    Moreover, LC-based multilayers offer additional advantages, such as enhanced flexibility and bending that makes them promising candidates for tunable X-ray optical components.

    \begin{figure}[t!]
        \center
        \includegraphics{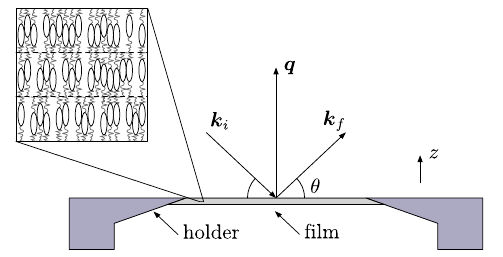}
        \caption{A free-standing smectic film  is drawn across the opening in a holder. The inset shows the cross-section of the film, with the molecules of the liquid crystal arranged in a stack of smectic layers. Here, $\bm{k}_i$ and $\bm{k}_f$ are the incoming and outgoing wave vectors, respectively (the specular reflectivity geometry is shown), $\bm{q}$ is the scattering vector, and $\theta$ is the angle of scattering.}
        \label{fig:standing_smectic_film}
    \end{figure}

    The simplest layered phase among LCs is smectic-A (Sm-A), in which the rod-like molecules are organised in stacks of liquid layers with the long molecular axes parallel to the layer normal, Fig.~\ref{fig:standing_smectic_film}. 
    Hence, a periodic structure exists in one dimension: the elongated molecules form a density wave along the layer normal, while the system remains fluid in the plane of layers. 
    When stretched on a frame, smectic LCs, due to their layered structure, form free-standing smectic films (FSSF) in which the smectic layers align parallel to the two air-film interfaces, Fig.~\ref{fig:standing_smectic_film}. 
    The FSSF are flat because the surface tension minimises the surface area of the film. Apart from the edges such films can be considered as substrate-free. 
    The alignment of the smectic layers is almost perfect, allowing studying of the single-domain samples of various thicknesses. 
    The surface area can be as large as a few $\text{cm}^2$, while the thickness can be varied up to thousands of layers (units and tens of microns).

    The FSSF have been extensively studied starting from the 1970s (for reviews, see~\cite{pieranski1993membranes, oswald2005crystals}).
    The structure of FSSF has been thoroughly investigated by specular and diffuse x-ray reflectivity~\cite{dejeu2003membranes}.
    These studies have led to significant new results concerning the growth of the mean square layer displacements with the sample size and the development of interlayer structure as a function of the film thickness.
    The Sm-A phase represents a classical example of a system with a low-dimensional order: it possesses only one direction of translational order in a three dimensional (3D) medium. 
    As a result long-range positional order is destroyed by thermal fluctuations of the system: the mean-square layer displacements diverge logarithmically with the sample size (Landau-Peierls instability)~\cite{landau1980statistical}. 
    A similar divergence due to long-wavelength displacement modes makes 2D crystals unstable. Fortunately, above logarithmic divergence is slow and allows the smectic samples of macroscopic size to exist~\cite{degennes993crystals}.  
    
    An alternative way to create well aligned smectic films is to deposit them on the solid substrate. 
    The films can be prepared by conventional methods like spin coating~\cite{olbrich1993smectic} or the Langmuir-Blodgett technique~\cite{geer1995polymer}. 
    In both cases the smectic film will be aligned due to anchoring forces at the film-air and film-substrate interfaces. 
    However, smectic films on a solid substrate suffer from two major drawbacks. First, such films are often unstable with respect to dewetting, like isotropic liquid films~\cite{herminghaus1998metal}. 
    The dewetting proceeds via an increase of the average roughness at the free surface, leading at a certain stage to the formation of the large-size holes. 
    Second, the static substrate roughness easily propagates into the film. 
    Because of the small compressibility of the system the characteristic length of the exponential decay of the static layer undulations can be as large as tens of microns~\cite{dejeu2003membranes, degennes993crystals}.
        
    For this reason, we investigated here the properties of Bragg mirrors based on freely suspended LC films. 
    Previously, we conducted synchrotron reflectivity studies of the FSSF of liquid crystal material 4O.8~\cite{fera1999films, fera2001scattering}.
    Here we analyzed the reflectivity spectrum of a 4O.8 film composed of approximately 80 smectic layers. 
    Our results demonstrated that the entire reflection curve, including the Bragg peaks from the multilayer smectic structure, could be well fitted within the model based on the kinematical theory of X-ray diffraction
    (see Section 3.1, below).
    This indicates that the film thickness (about $220$ nm) is too small to show the features predicted by the dynamical theory of X-ray diffraction. 
      
    The kinematical theory of X-ray diffraction, or the first Born approximation, assumes that multiple scattering effects at lattice planes are negligible and can be disregarded.
    This approximation is valid for relatively thin multilayer structures. Contrary to this, the dynamical theory accounts for multiple reflections at the interfaces between layers, and the resulting interference effects~\cite{pietsch2004high}.
    However, the transition from kinematical to dynamical regime of X-ray diffraction in multilayer systems can hardly be formulated on the quantitative level. 
    Authier and  Malgrange~\cite{authier1998diffraction} analysed this transition based on the concept of the extinction length.
    The latter is defined for a given Bragg peak as the depth at which the transmitted beam is completely diffracted due to multiple reflections at the internal interfaces measured within the spectral bandwidth of the Bragg reflection.
    According to their study, for the kinematical approximation to remain valid, the multilayer thickness must be less than one-tenth of the extinction distance. 
    This implies that for a given multilayer thickness, the longer the extinction distance, the more accurate the kinematical approximation.  Another attempt to estimate the crossover thickness was proposed by Muniz et al.~\cite{muniz2016scherrer}, 
    who compared the width of the Bragg peak obtained from Scherrer equation (calculated within the kinematical approach) with the peak width derived from the dynamical theory.
    However, it is important to note that such estimates remain semi-analytic in nature, as they are done numerically.
    
    In this study, we investigate the transition from the kinematical to the dynamical regime of X-ray diffraction using the multilayer free-standing smectic film as a model object. 
    We adopt an idea from~\cite{muniz2016scherrer} where the peak widths obtained by kinematical and dynamical theories are compared.
    Furthermore, based on this idea, we derive an analytical criterion for the crossover at which the kinematical regime of diffraction changes to the dynamical regime. 
    To the best of our knowledge, such an analytical approach has not yet been described in the relevant literature.
    Using this criterion, we analyze how different parameters of the film structure affect the margin between the kinematical and dynamical regimes. 
    This approach enables the determination of the crossover thickness (i.e., the critical number of layers) at the threshold between these regimes. 
    To support our results, we simulate the X-ray reflection of a real LC sample and compare our calculations with synchrotron data.

\section{X-ray reflectivity of a smectic multilayer: experiment and analysis}

    \begin{figure}[b!]
        \center
        \includegraphics{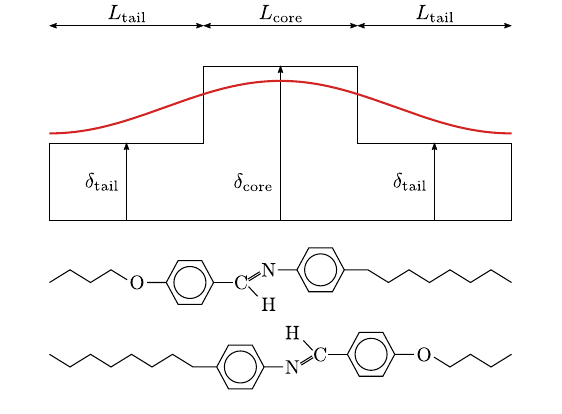}
        \caption{The two-step box model is used to represent the electron density of a single layer. The smectic periodicity is formed by two 4O.8 molecules with opposite up-down orientations. $\delta_{\text{core}}$, $\delta_{\text{tail}}$, $L_{\text{core}}$, $L_{\text{tail}}$ define the electron density and the length of the molecular fragments within the two-step box model, respectively. The red line illustrates the result of convoluting the box model with a Gaussian kernel, which is performed to account for layer displacement fluctuations.}
        \label{fig:box_like}
    \end{figure}

    The compound investigated, N-4-n-butoxybenzilidene-4-n-octylaniline, abbreviated as 4O.8, was obtained from Frinton Laboratories, Inc.
    The X-ray measurements (XRR) were performed within the temperature range of the Sm-A phase (around $60$°C).
    Freely suspended smectic films were drawn across a $10 \times 23$ $\text{mm}^2$ hole in a stainless steel frame and placed in a temperature controlled two-stage oven evacuated to $< 10^3$ Pa in order to reduce air scattering (for details see~\cite{fera2001scattering}).
    X-ray reflectivity measurements were carried out at beamline BW2 of HASYLAB (DESY, Hamburg) at an energy of $10.0$ keV (the wavelength $\lambda = 0.124$ nm).

    In a specular reflectivity experiment an incident beam of wave number $k = 2 \pi / \lambda$ is reflected at an interface.
    The incident wave vector $\bm{k}_i$ , the reflected wave vector $\bm{k}_f$ and the surface normal lie in the same (scattering) plane.
    Consequently, the wave-vector transfer $\bm{q} = \bm{k}_f - \bm{k}_i$ is parallel to the surface normal. Its projection on the z-axis is $q_z = 2k \sin \theta$ ($z$ is a coordinate axis along the surface normal and $\theta$ is an incoming angle), see Fig.\ref{fig:standing_smectic_film}.
    The quality of the films was checked by measuring rocking curves at different $q_z$.
    The mosaicity of the films was determined to be typically $0.005^\circ$ full width at half maximum FWHM.
    In the finite size smectic films the internal periodic structure generates the broadened Bragg peaks centred at $q_z = 2\pi m/D$, where $D$ is the layer spacing and $m$ is an integer.
    For 4O.8 films the  position of the first Bragg peak from the smectic layering corresponds to $q_z = 2.21 \text{nm}^{-1}$ ($D = 2.83$ nm).
    In the case of a FSSF, reflection also occurs at the opposite interface, leading to constructive or destructive interference in dependence of the incoming angle (Kiessig fringes).
    The period of the Kiessig fringes is inversely proportional to the film thickness $L$. Hence the number of smectic layers in the film $N = L/D$ is immediately determined from the specular diffraction profile.
    
    The XRR data were fitted to a theoretical model.
    As a forward model, the kinematical theory was used. Its formulation for the planar multilayer system is well known~\cite{pershan1984reflectivity,als-nielsen1985liquid}.
    This approach is also briefly described in the next section for reference.
    The slab model for the electron density profiles was used to parameterize the structure.
    At first the form factor of the orthogonal smectic layer was built up, which corresponds to two 4O.8 molecules with opposite up-down orientations, placed together to fit  the smectic layer periodicity.
    As a result a boxlike function shown in Fig.~\ref{fig:box_like} was obtained.
    To take the layer displacement fluctuations into account smectic layers were approximated by a convolution of a box function with a Gaussian of width $\sigma$~\cite{dejeu2003membranes, fera1999films}.
    The fitting parameters in this model include the ratios $\delta_{\text{core}}/\delta_{\text{tail}}$, $L_{\text{core}} / L_{\text{tail}}$ and the Gaussian spreading parameter $\sigma$, where $\delta_{\text{core}} / \delta_{\text{tail}}$ and $L_{\text{core}} / L_{\text{tail}}$ represent the contrast in electron density within the layer and the difference in length of the cores and tails of the paired molecules, respectively.
    The first two parameters are the same for all of the layers in the stack, while parameter $\sigma$ varies continuously from the beginning to the end of the multilayer structure. 
    We also added an additional parameter to our model that takes into account a small area with enhanced electron density of $s_{\rm spacing}$ at the border between the model layers. This area is smoothed as a result of the convolution.
    This enhanced electron density is due to the specific locations of the end-groups of the hydrocarbon chains within the LC molecules, see Fig.~\ref{fig:box_like}.

    \begin{figure}[t]
        \center
        \includegraphics{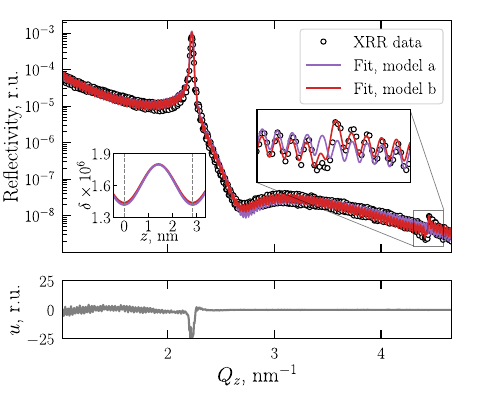}
        \caption{The best-fit of the XRR data using different models of electron density distrbution across the smectic layer stack.
        Model (a) corresponds to fit made with three independent parameters: $\delta_{\text{core}}/\delta_{\text{tail}}$, $L_{\text{core}} / L_{\text{tail}}$ and $\sigma$ (violet curve), while model (b) contains an additional delta-like function that
        describes a small area with enhanced
        electron density between the layers in the model,
        as discussed in the text (red curve).
        Both models provide a good fit along the whole reflectivity curve, as well as around the 1-st Bragg peak -- see inset in the middle-left of the graph, showing the electron density distribution within the single smectic layer, where the vertical dashed lines indicate the boundaries of one period.
        However, model (a) fails to reproduce the 2-nd order Bragg peak, while model (b) makes this perfectly -- see inset in the up-right of the graph.
        The lower part of the graph shows the discrepancy $u$ between model b and the experimental data.}
        \label{fig:fit}
    \end{figure}

    The results of the fitting of the experimental reflectivity data for the 4O.8 free standing film containing about 80 smectic layers are shown in Fig.~\ref{fig:fit}. 
    From the data fitting the $\delta_{\text{core}}$ parameter yielded a value of $2 \cdot 10^{-6}$, which is consistent with the known range of $\delta$ for liquid crystals, typically between $2$ and $3$ in units of $10^{-6}$.
    The ratio $\delta_{\text{core}} / \delta_{\text{tail}}$ was found to be $1.65$, aligning closely with ~\cite{dejeu2003membranes, fera1999films}.
    The values of $L_{\text{core}}$ and $L_{\text{tail}}$ were determined to be $1.42$ nm and $0.705$ nm, respectively, resulting in a total molecular length of $2.83$ nm, which agrees well with established data for 4.O8.
    The parameter $s_{\rm spacing}$ was found to be $0.05$~nm~$\times$ $1.8 \cdot10^{-6}$.
    According to our fitting this parameter must be taken into account
    to reproduce correctly the second Bragg peak from the multilayer structure, see Fig.~\ref{fig:fit}.

    \begin{figure}[t]
        \center
        \includegraphics{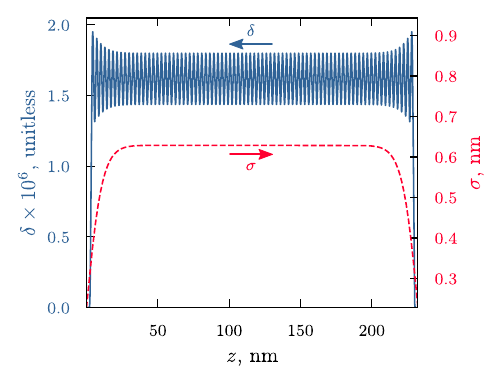}
        \caption{Electron density $\delta$-profile across the smectic film (solid blue line),
                  obtained by convoluting a set of box-like functions with a Gaussian kernel of varying roughness $\sigma$
                  (dashed red line). The layer displacement fluctuations are partly "frozen" at the FSSF borders due to interaction with the air-film interfaces in accordance with~\cite{dejeu2003membranes, fera1999films, fera2001scattering}.}
        \label{fig:profile}
    \end{figure}

    To simulate layer fluctuations, the initial approximation of the $\delta$-profile is convoluted with a Gaussian kernel.
    Such a procedure is the most common way to take fluctuations of layer displacements or dynamical roughness into account~\cite{dejeu2003membranes}.
    The fluctuation profile in the film together with the electron density $\delta$-profile is shown in Fig.~\ref{fig:profile}.
    The XRR curve, calculated on the basis of on the final $\delta$-profile, is depicted in Fig.~\ref{fig:fit}.
    The resulting quality of a fit is $\chi^2=5.2$. 

    Thus, a good agreement with the experimental data can be obtained for an 80-layer-thick smectic film using the kinematical approach. In this context, we would like to emphasize two points. First, the depth profile within the freely suspended smectic film is so pronounced that XRR becomes sensitive to subtle variations in the electron density near the internal layer interfaces. This enables the accurate modeling of the extremely weak second Bragg peak observed from the 4O.8 smectic film. Second, the experimental data exhibit no signatures of dynamical diffraction effects, such as Darwin plateau, which are characteristic of strong multiple scattering regimes. This observation immediately raises a question: is it possible to observe dynamical diffraction from smectic multilayers, and if so, what minimum layer stack thickness would be required for such effects to become visible? These questions will be addressed in the following sections. Prior to this, we briefly outline both the kinematical and dynamical approaches to calculating XRR of planar layered systems.

\section{Theory}

    The problem of calculating the X-ray reflectivity from a layered structure can be formulated as a Helmholtz equation:
    \begin{equation}
        (\Delta + k^2) E(\bm{r}) = V(z)E(\bm{r}) 
        .
        \label{eq:wave}
    \end{equation}
    Where $E$ is the electric field,
    $k$ is the wavenumber in vacuum,
    and $V$ is the scattering potential,
    which in the case of X-rays is $V(z) = -k^2\chi$ with dielectric susceptibility $\chi$.
    Since we are considering layered structures,
    we assume that $V$ depends only on the vertical coordinate $z$. Considering the relationship between the refractive index $n(z)$ and dielectric susceptibility $\chi$, the scattering potential $V(z)$ can be  expressed as $V(z) = k^2[1 - n^2(z)]$.
    For X-rays, the refractive index is written as 
    $n(z) = 1 - \delta(z) - i \beta(z)$,
    where $\delta(z)$ and $\beta(z)$ are the dispersion and absorption terms, respectively.
    We write Eq.~(\ref{eq:wave}) in the scalar approximation,
    since polarization effects can be neglected for grazing incidence in X-rays.
    XRR is calculated by solving this equation and calculating the far-field expansion components.
    It can be shown that for perfectly one-dimensional systems,
    such as a multilayer with no interfacial roughness, the dynamical theory happens to be exact.
    In the text below, we briefly describe the formulation of both theories used in this work.

\subsection{Kinematical approach to X-ray reflectivity}
    Within kinematical theory, the amplitude $r$ of a wave specularly reflected from a planar structure $V(z)$ can be calculated using a well-known~\cite{pershan1984reflectivity, als-nielsen1985liquid} formulae:
    \begin{equation}
        r(q_z) = -\frac{i}{q_z}\int_{-\infty}^{+\infty} V(z) e^{-i q_z z}dz
        ,
        \label{eq:kinematic_born}
    \end{equation}
    where $q_z = 2k \sin \theta$ is the momentum transfer and
    $\theta$ is the angle of incidence.
    We find the works\cite{zhou1995quantitative, zimmermann2005advanced} very instructive for a detailed derivation.
    This approach assumes that there is only a specularly reflected beam, while diffuse scattering is canceled out because the scattering potential is constant with respect to lateral translation, i.e.,
    the smoothness of the interfaces between the layers.
    The reflected intensity observed at the specular angle $\theta$
    is then proportional to $I = |r|^2$.
    It is easy to validate the convergence of the integral in Eq.~(\ref{eq:kinematic_born}) by integration by parts, which yields~\cite{als-nielsen1985liquid}:
    \begin{equation}
        r(q_z) = -\frac{1}{q^2_z} \int_{-\infty}^{+\infty} \frac{\partial V}{\partial z} e^{-i q_z z}dz
        .
        \label{eq:r_kinematical}
    \end{equation}

    It is convenient to interpret Eq.~(\ref{eq:r_kinematical}) in terms of the electron density $\rho$.
    By approximating $V(z)$ while neglecting second-order terms, one can write $V(z) \approx 2k^2[\delta(z) - i\beta(z)]$. Typically for X-rays and EUV, the absorption term $\beta$ is 2-3 orders of magnitude smaller than the dispersion term $\delta$, allowing us to neglect it,
    especially in our LC multilayer case.
    Then the scattering potential in terms of the electron density can be written as $V(z) \approx 4\pi r_{\rm e} \rho(z)$, where $r_{\rm e}$ is the classical electron radius.
    Thus, Eq.~(\ref{eq:r_kinematical}) implies that scattering occurs on the gradient of electron density, i.e. interfaces in a multilayer LC play the same role as crystallographic planes of an ordinary single crystal in X-ray diffraction. Furthermore, Eq.~(\ref{eq:r_kinematical}) implies that the reflected intensity decays as $I \sim q_z^{-4}$.

    Eq.~(\ref{eq:r_kinematical}) allows a direct calculation of the observed reflected intensity and the electron density distribution of a multilayer structure. 
    It is particularly effective in situations where multiple scattering events can be neglected.
    This makes it suitable for cases where the reflecticity is sufficiently low, in most practical cases that is $|R|<0.1$, as demonstrated in~\cite{kaganer1995bragg}.
    Thereby, it is not suitable for the total external reflection and high intensity diffraction peaks. 
    However, it still provides an accurate estimate of the intensity of higher order ($m > 1$) diffraction peaks, which are typically of lower intensity.
    In the specific context of LC multilayer systems, the kinematical theory adequately describes not only higher order diffraction peaks, but first order diffraction as well.
    This has been demonstrated earlier for smectic samples of various thicknesses~\cite{dejeu2003membranes, fera1999films, fera2001scattering, kaganer1991crystals}.
    We assume that this is due to the low electron density in LC compared to typical solid state Bragg mirrors.
    However, to strictly justify this observation, an exact calculation for the reflected intensity becomes necessary, which can be done through the dynamical theory.

\subsection{Dynamical approach to X-ray reflectivity}

    The dynamical theory, introduced by Darwin and later refined by Ewald and Laue, provides a framework that describes diffraction in crystals as the far-field representation of an electromagnetic standing wave formed between crystallographic planes when diffraction conditions are satisfied. It is generally written for 3D periodic structures such as perfect crystals. The theory accounts for multiple scattering events. However, this is done at the expense of simplifying the scattering potential. The scattering potential is approximated by a truncated Fourier series that typically includes only two to four components. Each component is related to each diffraction peak by Laue conditions.
    Therefore, the dynamical theory is valid only near the nodes of the reciprocal space considered in expanding $V$, i.e., around a limited number of preselected diffraction peaks.
    
    However, it is possible to implement a semi-analytical solution within the dynamical theory for 1D systems that is accurate both near and far from diffraction peaks.
    This is done by finding a solution to Eq.~(\ref{eq:wave}),
    assuming that $V={\rm const}$.
    Then, the structure is considered as a stratified medium divided into layers with constant scattering potential in each layer. Furthermore, the solution for each layer is constrained by applying boundary conditions.
    This was first introduced by Parratt, who formulated it in terms of recursive equations~\cite{parratt1954solids}.
    Alternatively, the same approach can be realized in terms of Abel\'{e}s matrices~\cite{abeles1950matrix}, or in terms of optical transfer matrices~\cite{gibaud2000scattering}.
    We use the latter formalism for the simulations.
    
    The solution of Eq.~(\ref{eq:wave}) for $j$-th layer with defined constants $\delta$ and $\beta$ is in the form of a standing wave:
    \begin{equation}
        E_j (\bm r) = (r_j e^{i k_{z,j} z} + t_j e^{-i k_{z,j} z} )e^{i\bm{k}_\parallel \cdot \bm{r}}
        ,
    \end{equation}
    where $r_j$ and $t_j$ denote the amplitudes of the forward- and backward-propagating components of the standing wave within the $j$-th layer, respectively.
    The reflection coefficient, $r(q_z)$, is proportional to the amplitude of the reflected wave at an infinite distance from the sample, such that $r(q_z) \propto r_0$.
    Under rigid boundary conditions ($t_0 = 1$ and $r_{N} = 0$): $r(q_z) \equiv r_0$. 
    The vertical component of the wave vector obeys spherical dispersion:
    $k_{z,j}^2 = (1+\chi)k^2 - k_\parallel^2$, or approximately
    $k_{z,j}^2  \approx k^2 (\sin^2\theta - 2\delta_j - 2i \beta_j)$. 
    Considering the boundary conditions for the electric field and its gradient,
    one can write 
    \begin{equation}
     \begin{bmatrix}
        r_0 \\
        t_0
        \end{bmatrix} =
        \mathbf{M}
        \begin{bmatrix}
        r_{N} \\
        t_{N}
        \end{bmatrix}
        ;
        \quad
        \mathbf{M}=
        \mathbf{R}_{0,1}
        \mathbf{T}_1
        \mathbf{R}_{1,2}
        \ldots
        \mathbf{R}_{N-1, N}
        .
        \label{eq:system}
    \end{equation}
    Here, $\mathbf M$ is the characteristic matrix of a multilayer structure.
    It relates the amplitude coefficients in the ambient area above the multilayer
    ($0$-th medium) and in the substrate
    ($N$-th medium).
    It is composed as a product of $\mathbf R$ and $\mathbf T$
    matrices defined for each medium.
    The matrix $\mathbf{R}_{j,j+1}$ represents the transformation of the amplitude due to the transition between the media $j$ and $j+1$:    
    \begin{equation}
        \mathbf{R} =
        \begin{bmatrix}
        m_+ & m_- \\
        m_- & m_+
        \end{bmatrix}
        ,
    \end{equation}
    where $m_\pm = (k_{z,j} \pm k_{z,j+1})/2k_{z,j}$ . 
    The matrix $\mathbf T$ propagates the amplitude through the $j$-th layer:
    \begin{equation}
        \mathbf{T} =
        \begin{bmatrix}
        e^{-i k_{z,j} h_j} & 0 \\
        0 & e^{i k_{z,j} h_j}
        \end{bmatrix}
        .
    \end{equation}
    By solving a simple $2 \times 2$ matrix equation~(\ref{eq:system})
    the reflected amplitude is calculated as
    $r_0 = M_{12}/M_{22}$.
    Thus, to calculate the X-ray reflectivity (XRR) curve, the process begins with defining the multilayer profile parameters, including the dispersion and absorption terms, and thicknesses of all layers in the system, as these govern the optical and structural properties of the multilayer system. 
    The next step is to calculate the wave vector components $k_{z,j}$ for each layer, taking into account $\delta(z)$, $\beta(z)$ and the incident angle  to characterize the wave propagation within the medium.
    Using this information, the transition matrices $\mathbf{R}$, which describe the boundary conditions at the interfaces between layers, are constructed. Additionally, the propagation matrices $\mathbf{T}$ are determined to describe the transmission and phase shift of the wave as it passes through each layer, incorporating the effects of layer thickness, absorption, and dispersion.
    Finally, the matrices $\mathbf{R}$ and $\mathbf{T}$ are sequentially multiplied for all layers, starting from the surface and progressing through the entire multilayer structure, to obtain the transfer matrix $\mathbf{M}$.
    The reflection coefficient $r$ is then extracted, enabling the determination of the reflectivity profile as a function of incident angle or momentum transfer.
    In this way, the procedure to calculate XRR with the dynamical theory is indirect and not immediately reversible, in contrast to the Fourier transform in kinematical theory.

\subsection{A criterion for kinematical to dynamical diffraction crossover}

    We have considered two theories for XRR.
    Kinematical theory is an approximation that conveniently relates the depth profile of the structure and the reflectivity as a Fourier pair.
    This is convenient because one can use properties of the Fourier transform for the calculation.
    For example, one can extract the depth profile directly~\cite{zimmerman2000reflection}, or use FFT algorithms for fast computation.
    As a rule of thumb, the kinematical theory applies to diffraction peaks with a reflectivity one or more orders of magnitude lower than the incident beam: $R \le 0.1$~\cite{kaganer1995bragg}.
    A more quantitative analysis is proposed in~\cite{muniz2016scherrer}.
    FWHM of the peak calculated numerically with kinematical (k) and dynamical (d) theories were compared to each other.
    These considerations were made in the context of X-ray powder diffraction.
    Thereby, in~\cite{muniz2016scherrer} FWHMs of diffraction peaks were calculated with respect to different sizes of the crystallites.
    At lower sizes the FWHMs are identical $\omega_{\rm k}/\omega_{\rm d}=1$.
    With the increase, at a certain crystallite size, the results differ in such a way that $\omega_{\rm k}/\omega_{\rm d}<1$.
    In kinematical theory, the width of the diffraction peak is inversely proportional to the size of a periodic structure.
    However, the dynamical theory suggests that the peak cannot be narrowed below the Darwin width, while the kinematical theory does not take this into account. 
    Physically, the Darwin width describes the flattening of the peak due to the formation of the standing wave, or similarly, the Borrmann effect~\cite{authier2006diffraction}.
    These are purely dynamical effects, i.e. related to the multiple scattering.
    Hence there is a divergence with kinematical theory for sufficiently thick multilayers.
    The number of periods for which this occurs represents the switch from the kinematical to the dynamical regime.

    Based on this consideration, a criterion for the critical number of periods $N_{\text{c}}$ at which the diffraction regime switches can be derived.
    Within the kinematical theory,
    the FWHM $\omega_{\text{k}}$ of the diffraction peak is described by the Scherrer equation~\cite{patterson1939scherrer}:
    \begin{equation}
        \omega_{\text{k}} = \frac{K \lambda}{DN\cos \theta_{\text{B}}}      
        .
        \label{}
    \end{equation}
    Here, $\theta_{\text{B}}$ is the Bragg angle,
    $D$ is the period,
    $\lambda$ is the beam wavelength
    and $N$ is the number of periods.
    $K$ is the calibration constant that accounts for the shape of the crystallite in the original formulation for the crystal powder.
    In effect, it is a correction for the deviation of the peak from the Gaussian shape.
    On the other hand, for the dynamical theory in the two-beam approximation, the Darwin width is given by~\cite{authier1998diffraction}
    \begin{equation}
        \omega_{\text{d}} = \frac{\text{Re}\{ \sqrt{ \chi_h \chi_{\overline h} } \} }{2 \sin \theta_{\text{B}}}      
        ,
        \label{}
    \end{equation}
    where $\chi_h$ is the Fourier component of the $\chi(r)$ calculated for the $m$-th order of diffraction i.e. at  $h= 2\pi m/D$:
    \begin{equation}
        \chi_h = \frac{1}{D} \int^D_0 \chi(z) e^{-ihz} dz    
        .
        \label{}
    \end{equation}
    Finally, by assuming $\omega_{\rm k} = \omega_{\rm d}$, and also assuming Friedel symmetry $\chi^*_h = \chi_{\bar{h}}$,
    we derive an expression for the critical number of periods: 
    \begin{equation}
        N_c = 2K \tan{\theta_{\rm B}} \dfrac{\lambda}{D} \dfrac{1}{|\chi_h|}
        ,
        \label{eq:N_c}
    \end{equation}
    Thus, the critical number of periods is defined by the scale of the diffraction problem, namely the ratio of wavelength to period $\lambda/D$, and is inversely proportional to the square root of the structure factor, i.e. $|\chi_h|$.
    Once the model of the structure $\chi(z)$ is defined,
    using Eq.~(\ref{eq:N_c}) one can estimate the number of periods $N_c$
    in which the crossover from the kinematical to the dynamical regime of diffraction occurs.
    
    For our periodic multilayer structure with a periodic two-step box model (see Fig.~\ref{fig:box_like}),
    the Fourier component $\chi_h$ can be easily calculated, also taking into account the roughness of the interfaces by convolution with a Gaussian profile. With this we can derive a version of Eq.~(\ref{eq:N_c}) specific to our model:
    \begin{equation}
        N_c =  \dfrac{Kh \lambda \tan{\theta_{\rm B}}}{2\Delta\chi \sin{(\pi m \Gamma)}}
        e^{h^2 \sigma^2/2} 
        ,
    \end{equation}
    where $\Delta \chi \approx 2|\delta_{\rm core} - \delta_{\rm tail}|$ is the optical contrast between layers in a period,
    $\Gamma$ is the two-step depth ratio in a layer: $\Gamma = d_{\text{core}}/D$
    and the last term in this equation arises from the Debye-Waller factor in which
    $\sigma$ is a mean roughness amplitude, 
    characterizing the thermal layer displacements in the film. 
    Thus, in terms of the structural parameters of the multilayer,
    the critical number of periods depends on the optical contrast, the depth ratio, and the roughness amplitude.

\section{Numerical simulations}

    In Fig.~\ref{fig:fit} above we have shown the fitting of the experimental data for 4O.8 film comprising about 80 smectic layers, based on the kinematical theory. 
    This result confirms that the first Bragg peak in a LC system of this type is accurately described using the kinematical theory. 
    However, experimental data for FSSF with a larger number of layers are unavailable, and a transition to the dynamical regime may only occur with a substantial increase in the number of layers. 
    To determine the critical $N_{\rm c}$ value (see Eq.~(\ref{eq:N_c})), characterizing the difference between the first Bragg peak descriptions in the kinematical and dynamical theories, corresponding numerical experiments were conducted. 
    Parameters derived from the fitting of experimental data for 4O.8 film of the 80 layers thick were utilized to construct $\delta$-profiles corresponding to varying numbers of layers in the smectic layer stack. 
    To simplify the numerical analysis, regions of sharp rise and fall of the $\sigma$ parameter were excluded (see Fig.~\ref{fig:profile}), retaining only the plateau region. 
    Consequently, the roughness parameter $\sigma$ was assumed constant for each layer. 
    Specific parameters were varied, and the intensity in the region of the first Bragg peak was calculated using both the kinematical and dynamical theories. 
    Based on the obtained data, the full width at half maximum (FWHM) ratio was determined, with its deviation from unity indicating the onset of dynamical effects.

    \begin{figure}[t]
        \center
        \includegraphics{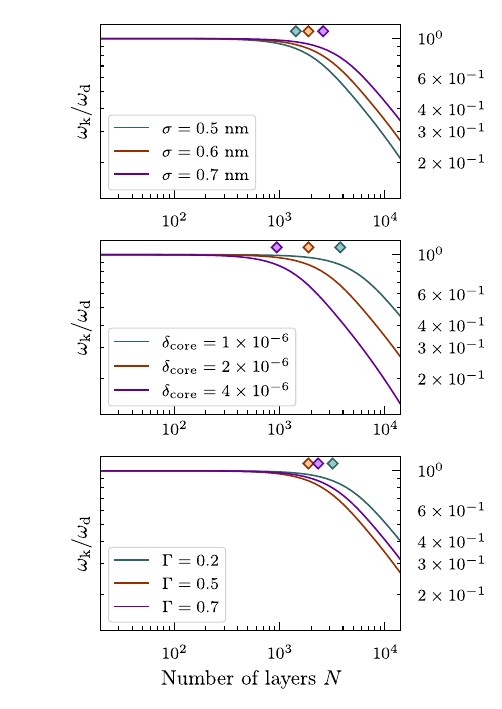}
        \caption{Dependence of the ratio of the FWHM of the first Bragg peak for kinematic and dynamic XRR curves on the number of layers $N$ in free standing smectic films. Each panel shows the effect of varying a single parameter while keeping the other two fixed: (top) thermal fluctuation profile $\sigma$, (middle) $\delta_{\rm core}$, and (bottom) two-step depth ratio: $\Gamma$. Diamond markers indicate a critical number of layers $N_{\rm c}$.}
        \label{fig:switch}
    \end{figure}
    \begin{figure}[t]
        \center
        \includegraphics{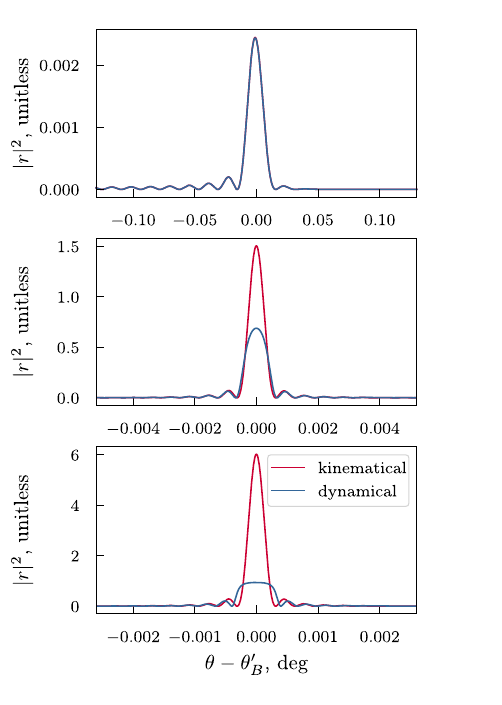}
        \caption{First Bragg peak of X-ray reflectivity calculated using kinematic and dynamical theories. The panels correspond to different numbers of layers in the film:(top) 80 layers, (middle) 2000 layers, and (bottom) 4000 layers. The calculations illustrate the formation of the Darwin plateau in the dynamical regime. }
        \label{fig:darwin}
    \end{figure}

    Figure~\ref{fig:switch} presents panels, each corresponding to the variation of one of three parameters, characterising the structure of the smectic film. 
    The top panel illustrates the dependence of the critical number of layers on the roughness parameter $\sigma$. 
    The middle panel reflects changes in the scale of the $\delta$-profile (corresponding to the electron density scale), while the bottom panel addresses variations in the two-step depth ratio $\Gamma$. 
    These findings substantiate the theoretical model developed to identify the transition area from the kinematical to the dynamical regime in describing the first Bragg peak of the LC structure. The critical point $N_{\rm c}$ is of the order of $10^3$ layers, indicating that a significantly large number of layers is required for dynamical effects to become well pronounced. Beyond this point the kinematical theory ceases to accurately describe the first Bragg peak of the LC structure. 
    The formation of the Darwin plateau is illustrated in Fig.~\ref{fig:darwin}.
    Here is a series of Bragg peaks calculated with both kinematical and dynamical theory for different numbers of periods. The Bragg peaks are calculated at about $\theta_{\rm B}'$ -- a modified Bragg angle to account for refraction within the structure. One can calculate $\theta_{\rm B}'$ from:
    $
        \cos^2\theta_{\rm B}' = 1+\chi_0 -{m^2\lambda^2}/{4D^2},
    $
    where $\chi_0$ is the 0-th Fourier component of $\chi(z)$,
    i.e., the average dielectric susceptibility of the structure.
    In Fig.~\ref{fig:darwin} the Darwin plateau is clearly present for the LC film comprising $4000$ smectic layers. Taking into account that the layer periodicity of 4O.8 films is about $3$ nm, this corresponds to a film of $12$ micrometers thick. For such FSSF it is evident that the peak modeled by the kinematical theory becomes narrower than the Darwin plateau.

\section{Discussion}

    In the preceding sections, we considered two theoretical approaches to describe X-ray diffraction from smectic multilayers prepared as free-standing films.
    Kinematical theory, which assumes negligible multiple scattering effects at lattice planes, is applicable to relatively thin films, where the width of the diffraction peak is inversely proportional to the size of the periodic structure.
    However, as the number of layers increases, dynamical effects start to take over, leading to the flattening of the Bragg peak, known as the Darwin plateau.
    In contrast, the dynamical matrix formalism explicitly accounts for multiple reflections at the interfaces between layers and the resulting interference effects.
    This approach enables a more accurate description of the Bragg peak profile and provides critical insight into the crossover between the kinematical and dynamical diffraction regimes.
    As a result, we have identified the threshold number of smectic layers beyond which the kinematical theory ceases to adequately model the first Bragg peak.
    
    In principle, the crossover film thickness can be estimated by analyzing the features of diffuse (nonspecular) X-ray scattering.
    It is well known that diffusely scattered waves exhibit maxima when the incidence and/or exit angles coincide with the critical angle $\theta_{\text{c}}$ for total external reflection -- so called Yoneda peaks~\cite{hol1994multilayers}.
    Scattering experiments on relatively thin free-standing smectic films systematically reveal the absence of Yoneda peaks in diffuse X-ray scans recorded at various $q_z$ values along the specular reflectivity curve~\cite{dejeu2003membranes}.
    In contrast, multilayered Bragg mirrors composed of inorganic materials exhibit strongly pronounced Yoneda peaks
    (see for instance data in~\cite{bruson1989yoneda}).
    This distinction is not accidental, but rather reflects the inherently dynamical nature of Yoneda peaks, which arise due to multiple scattering processes. 
    Specifically, these peaks can be quantitatively described using the distorted-wave Born approximation (DWBA)~\cite{hol1994multilayers}.
    The emergence of Yoneda peaks in diffuse scattering calculations for smectic films of a certain thickness might serve as an independent indicator of the transition to the dynamical diffraction regime.
    However, such calculations are computationally demanding and require specialized extensions of the dynamical matrix formalism.
    A detailed analysis of this problem falls beyond the scope of the present study.
    Nevertheless, we predict that Yoneda peaks should necessarily appear in diffuse X-ray scans of really thick (about $10$ microns) FSSF of Sm-A liquid crystal materials.
    
    In most thermotropic low-molecular-mass smectics only the first Bragg peak is clearly present, the intensity of the higher order diffraction being orders of magnitude less, see Fig.~\ref{fig:fit}.
    In contrast, polymeric smectics with mesogenic side chains~\cite{nachaliel1991crystal, davidson1992mesorphic} and smectic elastomers~\cite{obraztsov2008elastomers} are less compressible and thus higher-order harmonics are more pronounced.
    Furthermore, in such free-standing smectic films, the amplitudes of layer displacement fluctuations are approximately twice as small as those observed in low-molecular-mass smectics. The intensity of the successive harmonics is also influenced by the wave-vector dependence of the molecular form factor $F(q)$.
    For instance, $F(q)$ favors more intense higher harmonics for LC molecules with perfluorinated or siloxane terminal groups possessing higher electron density~\cite{ostrovskii1999lc}.
    All together enhances the potential of the above types of smectic ordering for the fabrication of multilayered Bragg mirrors.

    Polymeric side-chain smectic liquid crystals can be prepared as FSSF~\cite{link2000crystal}.
    By incorporating a crosslinking agent and subsequently exposing the material to UV radiation, a perfectly aligned smectic multilayer system with the mechanical properties of liquid crystalline elastomers (soft rubber) can be obtained~\cite{dejeu2011elastomers, wang2022composites}.
    These films exhibit high mechanical stability, making them promising candidates for high-quality adaptive X-ray optical elements.

\section{Conclusion}

    The kinematical theory effectively describes X-ray reflectivity curve or, to be more precise, Bragg peaks in smectic liquid crystals due to the relatively small value of electron density $\delta$.
    Its simplicity provides a practical advantage over the dynamical theory for thinner multilayers.
    However, as the number of layers increases, dynamical effects such as the Darwin plateau emerge, marking the point where the kinematical theory becomes insufficient.
    At this stage, only the dynamical theory can accurately model the reflectivity behavior and describe the multilayer structure.
    This study highlights the necessity of transitioning to the dynamical framework for thicker smectic films while reinforcing the utility of kinematical theory for simpler, thinner systems.
    These insights guide the design and optimization of liquid crystal-based Bragg mirrors for advanced optical applications.

\section*{Funding information}

    The theoretical part of this work was carried out within the framework of the state assignment of the National Research Center "Kurchatov Institute".
    The experimental study was supported by Russian Science Foundation Grant No. 23-12-00200.

\section*{Acknowledgments}

    The authors would like to thank Wim de Jeu from the FOM-Institute AMOLF in Amsterdam under whose guidance the X-ray reflectivity synchrotron measurements of the free standing smectic films were made and Andrea Fera from the same institute for his participation in experiments and scientific contributions.
    Special thanks are due to Vladimir Kaganer (PDI, Berlin) for valuable discussions.

\section*{Appendix: Roughness as a continuous parameter over depth}

    To accurately account for real thermal fluctuations in the calculation, a convolution with a Gaussian kernel was applied to the $\delta$ profile. Furthermore, it was shown from thermodynamic considerations~\cite{fera1999films} that the r.m.s. roughness $\sigma$ varies with depth $z$.
    Therefore, the convolution procedure needs to be modified so that the r.m.s. roughness is a continuous parameter: $\sigma(z)$.
    The convolution procedure was modified to account for this.
    We used the following formulation of convolution in which
    $G(z,z') \neq G(|z-z'|)$:
    \begin{equation}
        \delta_{\rm conv}(z) = \int_{-\infty}^{+\infty} \delta_{\rm initial}(z')G(z, z')dz'  
        ,
        \label{}
    \end{equation}
    where $\delta_{\rm initial}(z)$ represents a set of box-like functions, each corresponding to a layer, and $G(z, z')$ is the Gaussian kernel defined as:
    \begin{equation}
        G(z, z') = \frac{1}{\sqrt{2 \pi} \sigma (z')} \exp \left [ -\frac{(z-z')^2}{2 \sigma^2 (z')} \right ] 
        ,
        \label{}
    \end{equation}
    On a discrete grid, an alternative approach is to further reformulate the convolution as a matrix multiplication, simplifying the computational process.
    In this approach, the $\delta$-profile: $\delta(z)$ is represented as a sequence of discrete values that are combined into a vector $\bm{\delta}$.
    The convolution of the profile can then be expressed as matrix product:
    \begin{equation}
        \bm{\delta}_{\rm conv} = \bm{G} \bm{\delta}_{\rm initial}
        ,
        \label{}
    \end{equation}
    where the element of the kernel matrix $\bm G$ is constructed as:
    \begin{equation}
        G_{ij} = \dfrac{
                \dfrac{1}{\sqrt{2 \pi} \sigma_j} 
                \exp 
                \left [ -\dfrac{(z_i-z_j)^2}{2 \sigma_j^2} 
                \right ]}
                {\sum_k \dfrac{1}{\sqrt{2 \pi} \sigma_k} 
                \exp \left [ -\dfrac{(z_i-z_k)^2}{2 \sigma_k^2} \right ]}
        .
        \label{}
    \end{equation}
    The denominator term in the kernel ensures the preservation of the total integral of the resulting profile, maintaining the overall area under the box-like functions.
    The parameter $\sigma_i$  defines the fluctuation scale at depth $z_i$, and it was specifically chosen to vary continuously rather than in discrete steps from layer to layer.

    To model the shape of the fluctuation curve, a super-Gaussian function was selected, following the profile described in~\cite{dejeu2003membranes}.
    This function enables accurate representation of thermal fluctuations across the entire multilayer structure, particularly maintaining nearly constant values in the middle layers (see dashed red line in Fig~\ref{fig:profile}):
    \begin{equation}
        \sigma(z) = \dfrac{a}{\sqrt{2 \pi} s} 
            \exp
            \left[
            { -\dfrac{1}{2} \dfrac{|z-z_0|^p}{s^p}}
            \right]
        .
        \label{}
    \end{equation}
    The normalization factor was determined to be $a = 176.04$, width $s = 111.7$,  the center of the fluctuation curve set at $z_0= 116.2$~nm (half of the smectic film thickness), and $p = 18.3$.
    
    After computing the convolution $\delta_{\text{conv}}$, the obtained profile is used to calculate the X-ray Reflectivity (XRR) curve using both kinematical and dynamical theories for comparison.
    To account for inherent uncertainty in photon detection, the XRR curve undergoes convolution with a Gaussian kernel characterized by a standard deviation $\sigma_{\text{core}} = 6.8 \cdot 10^{-5}$.

\bibliography{bibliography}{}
\bibliographystyle{ieeetr}

\end{document}